\documentclass[aps,prl,twocolumn,groupedaddress,showpacs,floatfix]{revtex4-1}%
\usepackage{graphicx,amsmath,amssymb}
\usepackage{amsmath}
\usepackage{amsfonts}
\usepackage{amssymb}
\usepackage{graphicx}%
\providecommand{\U}[1]{\protect\rule{.1in}{.1in}}
\begin{document}
\title{Quantum Energy Teleportation without Limit of Distance}
\author{Masahiro Hotta}
\email{hotta@tuhep.phys.tohoku.ac.jp}
\affiliation{Department of Physics, Tohoku University, Sendai 980-8578, Japan}
\author{Jiro Matsumoto}
\email{jmatsumoto@tuhep.phys.tohoku.ac.jp}
\affiliation{Department of Physics, Tohoku University, Sendai 980-8578, Japan}
\author{Go Yusa}
\email{yusa@m.tohoku.ac.jp}
\affiliation{Department of Physics, Tohoku University, Sendai 980-8578, Japan}
\date{\today
}

\begin{abstract}
Quantum energy teleportation (QET) is, from the operational viewpoint of
distant protocol users, energy transportation via local operations and
classical communication. QET has various links to fundamental research fields
including black hole physics, the quantum theory of Maxwell's demon, and
quantum entanglement in condensed matter physics. However, the energy that has
been extracted using a previous QET protocol is limited by the distance
between two protocol users; the upper bound of the energy being inversely
proportional to the distance. In this letter, we prove that introducing
squeezed vacuum states with local vacuum regions between the two protocol
users overcomes this limitation, allowing energy teleportation over practical distances.

\end{abstract}

\pacs{03.67.Ac}
\maketitle

\section{Introduction}

Quantum fields in the vacuum state accompany spatially entangled energy
density fluctuations via the noncommutativity of energy density operators.
However, the eigenvalue of the total Hamiltonian can be set to zero by
discarding the zero-point energy, primarily because this energy exhibits the
fundamental property known as passivity \cite{passivity} and is of little use;
any intended local operation for extracting the zero-point energy out of a
field actually injects energy and excites the vacuum. The zero-point energy,
however, can be glimpsed through a spatial region with negative energy density
\cite{BD}, along with the Casimir effect \cite{C} and the Unruh effect
\cite{UW}. In such a spatial region, the quantum field can afford to attain a
lower energy density than the zero value of the vacuum state because it saves
the zero-point energy in the vacuum state. Of course, we have another region
with sufficient positive energy to ensure that the total energy is greater
than zero.

Recently, quantum information theory has revealed some exotic aspects of the
entangled energy density fluctuation of many-body systems in the ground state,
and it has been proven that quantum energy teleportation (QET)\ is possible
\cite{QET,bh}. QET is, from the operational viewpoint of distant protocol
users, energy transportation via local operations and classical communication
(LOCC). In contrast to the standard protocols of quantum teleportation
\cite{qt}, QET protocols involve energy transportation. QET is related to the
quantum theory of Maxwell's demon at low temperatures \cite{fgh} and the local
cooling problem of many-body quantum systems \cite{sc}. Furthermore, QET
provides insight into the black hole entropy problem \cite{bh}. QET can be
implemented in various research fields including spin chains \cite{sc},
harmonic chains \cite{hc}, and cold trapped ions \cite{coti}. Although QET has
not yet been experimentally verified, a realistic experiment was recently
proposed \cite{yih}; the experiment uses 1+1-dimensional chiral massless boson
fields of quantum Hall edge currents \cite{yoshioka}, and the teleported
energy may be expected to take values near $O(100)~\mu$eV with present technology.

\begin{figure}[ptb]
\includegraphics[clip]{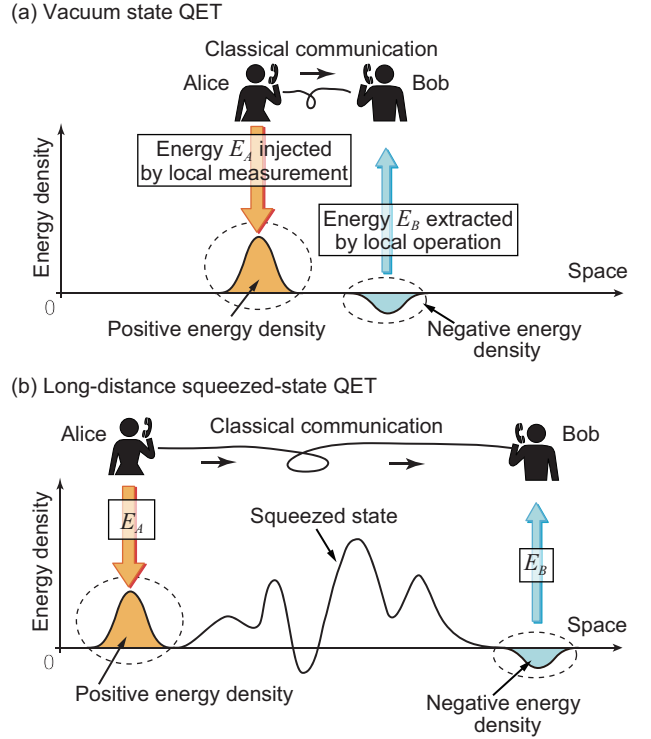} \caption{(Color online) Schematic diagram of
(a) vacuum state quantum energy teleportation (QET) protocol and (b)
long-distance squeezed-state QET.}%
\label{fig:modelQET}%
\end{figure}

Despite such experimental proposals for various physical systems
\cite{sc,hc,coti,yih}, a strong distance limitation has hampered experimental
verification; in QET over a distance $L$, the transferred energy $E_{B}$ of
1+1 dimensional massless scalar fields is bounded by
\begin{equation}
E_{B}\leq\frac{1}{12\pi L}, \label{1}%
\end{equation}
as long as vacuum-state QET protocols are adopted. Here, the natural unit
$c=\hbar=1$ is adopted. QET thus sends only a small amount of energy over a
long distance $L$. This vacuum-state QET distance bound appears for two
reasons: From an informational viewpoint, the spatial correlations of the
zero-point fluctuation, including quantum entanglement, decay as the distance
becomes large, and hence the amount of information for distant control of a
quantum fluctuation becomes small and only weak strategies for extracting
energy out of the distant zero-point fluctuation are available. From a
physical viewpoint, the localized negative energy induced by a QET protocol
cannot be separated from the positive energy injected by the measurement
device; otherwise, the negative energy excitation would exist without any
positive energy excitations. To make the total energy of the field
nonnegative, the negative energy excitation requires a sufficient amount of
positive energy at a close distance. However, there is an interesting
possibility that avoids this bound by using an exotic quantum state for QET,
as we argue in this paper.

We propose here a new version of the QET protocols that use a squeezed vacuum
state in place of the vacuum state between two protocol users (Fig.~1). In our
proposal, the spatial correlation of the quantum fluctuations with zero energy
is maintained even if the distance between the sender and receiver of QET is
very large. The negative energy induced by the extraction of positive energy
via QET is sustained not by the positive energy injected by the measurement
but instead by the excitation energy in the squeezed region of the state.
Thus, the bound in Eq.~(\ref{1}) is overcome and long-distance energy
teleportation can be achieved. The protocol may be implemented by adopting a
spatial expansion method, which is one strategy for realizing a long-distance
correlation, thus facilitating experimental verification of QET and
potentially contributing to quantum device applications.

The paper is organized as follows. In Section 2, a brief review of
vacuum-state QET is provided, and in Section 3, the distance bound in
Eq.~(\ref{1}) is derived. In Section 4, using a squeezed state, we outline the
new protocol that realizes QET without the limit of distance. The results are
summarized in Section 5.

\section{\bigskip Brief Review of Vacuum-State QET}

Recalling the experimental proposal using quantum Hall edge currents
\cite{yih}, let us consider a free massless scalar quantum field $\hat
{\varphi}$ in 1+1 dimensions that obeys the equation of motion
\[
\left(  \partial_{t}^{2}-\partial_{x}^{2}\right)  \hat{\varphi}=0.
\]
The general solution is obtained as the sum of a left-moving component
$\hat{\varphi}_{+}(x^{+})~$and a right-moving component $\hat{\varphi}%
_{-}(x^{-})~$using the light-cone coordinates $x^{\pm}=t\pm x$. For the
purposes of our discussion, we can focus solely on the left mover
$\hat{\varphi}_{+}(x^{+})$. In quantum Hall systems, $\hat{\varphi}_{+}%
(x^{+})$ describes the charge density fluctuation of the edge \cite{yoshioka}.
The energy flux operator is given by
\[
\hat{T}_{++}(x^{+})=:\hat{\Pi}_{+}(x^{+})^{2},
\]
where $\hat{\Pi}_{+}(x^{+})=\partial_{x^{+}}\hat{\varphi}_{+}(x^{+})$, and the
total energy operator is calculated as
\[
\hat{H}_{+}=\int_{-\infty}^{\infty}\hat{T}_{++}(x^{+})dx^{+},
\]
which is clearly a nonnegative operator. The vacuum state $|0\rangle$ is the
eigenstate with a zero eigenvalue of $\hat{H}_{+}$.

Let us briefly review vacuum-state QET using this $\hat{\varphi}~$\cite{bh}.
Consider two separate experimenters (say, Alice and Bob) who are able to
execute LOCC on this field in the vacuum state. Alice stays in the spatial
region $\left[  x_{1A},x_{2A}\right]  $ and Bob in $\left[  x_{1B}%
,x_{2B}\right]  $, with $x_{2A}<x_{1B}$. Bob's region is located to the right
of Alice's region, and the distance between them is $L=x_{1B}-x_{2A}$. Assume
that the initial state is the vacuum state $|0\rangle\langle0|$. The field
possesses zero-point fluctuation, and its nontrivial correlation is induced by
vacuum-state entanglement. Hence, if Alice obtains information about a local
fluctuation around her through a measurement, she simultaneously obtains some
information about a local fluctuation around Bob via the correlation. Although
the average value of the energy density in Bob's region remains zero after
Alice's measurement, Bob's local field in the post-measurement state carries
positive or negative energy, depending on Alice's measurement result. When the
result indicates the positive-energy case, Bob can extract energy from the
field after receiving the information from Alice. At time $t=0$, Alice
instantaneously conducts a general measurement \cite{nc} in $\left[
x_{1A},x_{2A}\right]  $. Although several measurements are useful for
realistic QET experiments \cite{m}, let us consider a simple measurement model
with a one-bit output, to grasp the essence of QET. Alice prepares a qubit $Q$
in $|-\rangle$, which is the down eigenstate of the third Pauli matrix
$\hat{\sigma}_{3Q}$, as a probe of $\hat{\Pi}_{+}(x^{+})$ in $|0\rangle$. The
interaction between the two states such that$\quad$%
\[
H_{int}(t)=\delta(t)\left(  \hat{A}-\frac{\pi}{4}\right)  \otimes\hat{\sigma
}_{2Q},
\]
where $\hat{\sigma}_{2Q}$ is the second Pauli matrix of $Q$, and $\hat{A}$ is
a local Hermitian operator defined as
\[
\hat{A}=\int_{-\infty}^{\infty}g_{A}(x)\hat{\Pi}_{+}(x)dx,
\]
with the real function $g_{A}(x)$ localized in $\left[  x_{1A},x_{2A}\right]
$. After switching off the interaction, a projection measurement of
$\hat{\sigma}_{3Q}$ is performed for $Q$, and Alice obtains a binary
measurement result of $\mu=0,1$ that corresponds to the eigenvalue $\left(
-1\right)  ^{\mu+1}$ of $\hat{\sigma}_{3Q}$. The (unnormalized)
post-measurement state of the field is given by $\hat{M}_{\mu A}|0\rangle$,
where $\hat{M}_{\mu A}$ are the measurement operators and are explicitly
computed \cite{bh} as
\begin{align}
\hat{M}_{0A}  &  =\cos\left(  \hat{A}-\frac{\pi}{4}\right)  ,\label{11}\\
\hat{M}_{1A}  &  =\sin\left(  \hat{A}-\frac{\pi}{4}\right)  . \label{12}%
\end{align}
A straightforward computation shows that the emergence probability of each
result $\mu$ is the same: $p_{\mu}=\langle0|\hat{\Xi}_{\mu A}|0\rangle=1/2$.
The post-measurement state for each $\mu$ is calculated as%

\[
\hat{\rho}_{\mu}=\frac{1}{p_{\mu}}\hat{M}_{\mu A}|0\rangle\langle0|\hat
{M}_{\mu A}^{\dag}.
\]
As a result of the passivity, this measurement injects an average energy of
\[
E_{A}=\sum_{\mu}p_{\mu}\operatorname*{Tr}\left[  \hat{H}_{+}\hat{\rho}_{\mu
}\right]  -\operatorname*{Tr}\left[  \hat{H}_{+}|0\rangle\langle0|\right]
\]
into the field and generates a positive energy wave packet. The injected
energy is computed as
\[
E_{A}=\int_{0}^{\infty}\left\vert \widetilde{g}_{A}(\omega)\right\vert
^{2}\omega^{2}\frac{d\omega}{4\pi}.
\]
Here, $\widetilde{g}_{A}(\omega)~$is the Fourier transform of $g_{A}(x)$.
Because $\hat{M}_{\mu A}$ includes only the left-mover operator $\hat{\Pi}%
_{+}(x)$, the wave packet moves to the left, i.e., further away from Bob.
Thus, Bob cannot receive the energy through direct propagation of the wave
packet. Note that the field in Bob's region is in a local vacuum state with
zero energy, although a wave packet with $E_{A}$ exists far from Bob. Assuming
that Bob receives the information of $\mu$ from Alice at time $t=T$, Bob then
performs an instantaneous unitary operation dependent on $\mu$ given by
\begin{equation}
\hat{U}_{\mu B}=\exp\left(  i\theta_{\mu}\hat{B}\right)  , \label{13}%
\end{equation}
where $\theta_{\mu}$ is a $\mu$-dependent real parameter fixed so as to
maximize the amount of teleported energy \cite{bh}, and
\[
\hat{B}=\int_{-\infty}^{\infty}g_{B}(x)\hat{\Pi}_{+}(x)dx
\]
with the real function $g_{B}(x)$ localized in $\left[  x_{1B},x_{2B}\right]
$. The post-operation state is computed as
\begin{align*}
\hat{\rho}_{QET}  &  =\sum_{\mu}\hat{U}_{\mu B}\exp\left(  -iT\hat{H}%
_{+}\right)  \hat{M}_{\mu A}|0\rangle\langle0|\hat{M}_{\mu A}^{\dag}\\
&  \times\exp\left(  iT\hat{H}_{+}\right)  \hat{U}_{\mu B}^{\dag}.
\end{align*}
It can then be verified that the total energy decreases during this local
operation by Bob. This implies that a positive amount of teleported energy
$E_{B}$ is extracted from the field in the local vacuum state as negative work
by Bob's operation:
\begin{align}
E_{B}  &  =\operatorname*{Tr}\left[  \hat{H}_{+}\sum_{\mu}p_{\mu}\exp\left(
-iT\hat{H}_{+}\right)  \hat{\rho}_{\mu}\exp\left(  iT\hat{H}_{+}\right)
\right] \nonumber\\
&  -\operatorname*{Tr}\left[  \hat{H}_{+}\hat{\rho}_{QET}\right]  >0.\nonumber
\end{align}
Simultaneously, a wave packet with negative energy $-E_{B}$ is generated in
Bob's region and begins to move to the left. It is useful at this point to
define the two-point correlation function $C_{\mu AB}$ as
\begin{equation}
C_{\mu AB}=\langle0|\hat{\Xi}_{\mu A}\hat{B}^{\prime}(T)|0\rangle/p_{\mu},
\label{20}%
\end{equation}
with
\[
\hat{\Xi}_{\mu A}=\hat{M}_{\mu A}^{\dag}\hat{M}_{\mu A}%
\]
and
\[
\hat{B}^{\prime}(T)=\int_{-\infty}^{\infty}g_{B}(x-T)\partial_{x}\hat{\Pi}%
_{+}(x)dx.
\]
Note that $C_{\mu AB}$ is real because of the operator locality: $\left[
\hat{\Xi}_{\mu A},\hat{B}^{\prime}(T)\right]  =0$. To maximize the extracted
energy, we fix the real parameter $\theta_{\mu}$ as
\[
\theta_{\mu}=\frac{2C_{\mu AB}}{G_{B}},
\]
where $G_{B}=\int_{-\infty}^{\infty}\left(  \partial_{x}g_{B}(x)\right)
^{2}dx$. This implies that positive energy is extracted from the field in the
local vacuum state as negative work by Bob's operation. The teleported energy
$E_{B}$ can be evaluated as
\begin{equation}
E_{B}=\frac{1}{G_{B}}\sum_{\mu}p_{\mu}C_{\mu AB}^{2}. \label{19}%
\end{equation}
From the correlation function,
\begin{equation}
\langle0|\hat{\Pi}_{+}(x_{B})\hat{\Pi}_{+}(x_{A})|0\rangle=-\frac{1}%
{4\pi\left(  x_{B}-x_{A}-i\epsilon\right)  ^{2}}, \label{14}%
\end{equation}
we can evaluate $C_{\mu AB}$ in a straightforward manner, and $E_{B}$ can be
explicitly computed \cite{bh} as
\begin{equation}
E_{B}=\frac{\left(  \int_{-\infty}^{\infty}\int_{-\infty}^{\infty}\frac
{g_{B}(x_{B})g_{A}(x_{A})}{\left(  x_{B}-x_{A}+T\right)  ^{3}}dx_{B}%
dx_{A}\right)  ^{2}}{\pi^{2}\int_{-\infty}^{\infty}\left(  \partial_{x}%
g_{B}(x)\right)  ^{2}dx\exp\left(  \frac{1}{\pi}\int_{0}^{\infty}\left\vert
\widetilde{g}_{A}(\omega)\right\vert ^{2}\omega d\omega\right)  }. \label{3}%
\end{equation}
The teleported energy $E_{B}$ is not larger than $E_{A}$ because of the
nonnegative property of the total Hamiltonian.

\section{\bigskip Distance Bound for Vacuum-State QET}

For vacuum-state QET, when the distance $L(=x_{1B}-x_{2A})$ between Alice and
Bob increases, $E_{B}$ in Eq.~(\ref{3}) decreases as $E_{B}\varpropto1/L^{6}$.
This damping behavior can be slightly improved to $E_{B}\varpropto1/L^{4}$ by
replacing $\hat{B}$ in Eq.~(\ref{13}) with $\hat{B}=\int_{-\infty}^{\infty
}\tilde{g}_{B}(x)\hat{\varphi}(x)dx$. A natural question then arises: To what
extent does another QET protocol improve this long-distance behavior of
$E_{B}$? As mentioned above, a stringent bound on the long-distance damping of
$E_{B}$ is given by Eq.~(\ref{1}) for any vacuum-state QET. This bound is
based on Flanagan's theorem \cite{F}, which asserts the following. Consider a
nonnegative continuous function $\xi(x)$ with $\xi(x\rightarrow\pm\infty)=0$
and define a Hermitian operator as
\[
\hat{H}_{\xi}=\int_{-\infty}^{\infty}\xi(x)\hat{T}_{++}(x)dx.
\]
The inequality
\[
\operatorname*{Tr}\left[  \hat{H}_{\xi}\hat{\rho}\right]  \geq-\frac{1}{12\pi
}\int_{-\infty}^{\infty}\left(  \partial_{x}\sqrt{\xi(x)}\right)  ^{2}dx
\]
holds for an arbitrary state $\hat{\rho}$, and applying this theorem to
vacuum-state QET yields Eq.~(\ref{1}).

Let $\hat{\rho}_{QET}$ denote the post-operation state following an arbitrary
vacuum-state QET. Assume that a wave packet with negative energy $-E_{B}$ is
generated in $\left[  x_{1B},x_{2B}\right]  $. In the intermediate region
between Alice and Bob, $\left[  x_{2A},x_{1B}\right]  $, the average energy
density vanishes. In the region to the left of Alice, $(-\infty,x_{2A}]$, a
wave packet exists with positive energy $E_{A}$. Let us impose the values
$\xi(x)=0$ for $x\in(-\infty,x_{2A}]$ and $\xi(x)=1$ for $x\in\left[
x_{1B},x_{2B}\right]  $ on $\xi(x)$. In the region $[x_{2B},\infty)$, it is
sufficient to assume that $\xi(x)$ slowly decreases to $0$. Resultantly,
\[
\operatorname*{Tr}\left[  \hat{H}_{\xi}\hat{\rho}_{QET}\right]  =-E_{B}%
\]
for an arbitrary $\xi(x)$ satisfying the above conditions. Thus,
\[
E_{B}\leq\frac{1}{12\pi}\inf_{\xi(x)}\int_{-\infty}^{\infty}\left(
\partial_{x}\sqrt{\xi(x)}\right)  ^{2}dx
\]
must be satisfied. The infimum of the $\xi(x)$ satisfying the above boundary
conditions is then obtained, using a variation method, from the function
$\xi_{opt}(x)$ obeying
\[
\xi_{opt}(x)=\left(  x/L\right)  ^{2}%
\]
for $x\in\left[  x_{2A},x_{1B}\right]  $. This results in the inequality of
Eq.~(\ref{1}). Note that a spatial region with negative energy can appear only
when another region with sufficient positive energy exists. If an excitation
with a fixed negative energy could be separated from a positive-energy
excitation by an infinitely large distance, then the positive-energy
excitation at the spatial infinity will not influence the negative-energy
excitation due to the locality of quantum field theory. This leads to an
apparent contradiction in the nonnegativity of the total energy in a broad
region surrounding the negative-energy excitation. Thus, in vacuum-state QET,
the negative-energy excitation generated by Bob should be located in the
neighborhood of the positive-energy excitation generated by Alice. This fact
yields the distance bound of Eq.~(\ref{1}).

\section{\bigskip Long-Distance Squeezed-State QET}

A long-distance QET is expected to open new doors in the development of
quantum devices. Thus, it is important to raise the question, can the distance
bound in Eq.~(\ref{1}) be overcome by some means? Interestingly, a loophole
can be found. Here, we propose the use of a squeezed state between Alice and
Bob instead of the vacuum state (Fig.~1) to achieve a long-distance QET. For
simplicity, let us set
\[
x_{2A}=-\left(  L+T\right)  /2=-d
\]
and
\[
x_{1B}+T=\left(  L+T\right)  /2=d.
\]
Consider a non-decreasing $C^{1}$ function $f(x)$ such that
\begin{equation}
f(x)=x\,+l/2 \label{01}%
\end{equation}
for $x\leq-d$ and%
\begin{equation}
f(x)=x-l/2 \label{02}%
\end{equation}
for $x\geq d$. Here, $l$ is a length parameter, and because $\partial
_{x}f(x)\geq0$, the parameter $l$ satisfies
\[
l\leq L+T=2d.
\]
In the following analysis, any $f(x)$ satisfying these conditions can be
applied. A typical example of $f(x)$ that exactly satisfies these conditions
is provided as a $C^{1}$-class odd function under the $x\rightarrow-x$
transformation as follows. A coordinate value $\bar{x}$ ($0<\bar{x}<d$)~and a
positive parameter $\Lambda$ satisfying $0<\Lambda<\left[  2(d-\bar
{x})\right]  ^{-1}$ can be used to define $f(x)$ as
\begin{equation}
f(x)=\left(  1-2\Lambda\left(  d-\bar{x}\right)  \right)  x \label{03}%
\end{equation}
for $0\leq x\leq\bar{x}$ and
\begin{equation}
f(x)=x-l/2+\Lambda\left(  x-d\right)  ^{2} \label{04}%
\end{equation}
for $\bar{x}\leq x\leq d$.$\ $In this case, the shift is given by
\[
l=2\Lambda\left(  d^{2}-\bar{x}^{2}\right)  .
\]
When $\Lambda\rightarrow\left[  2(d-\bar{x})\right]  ^{-1}$ and $\bar
{x}\rightarrow d$, the parameter $l$ approaches its upper limit, $L+T$. From
any $f(x)$, it is possible to define a complete set of mode functions
$\left\{  v_{\omega}(x)\right\}  $ of $\hat{\varphi}_{+}(x^{+})$ using the
relation \cite{BD}$~$%
\begin{equation}
v_{\omega}(x)=\frac{1}{\sqrt{4\pi\omega}}\exp\left(  -i\omega f(x)\right)  .
\label{05}%
\end{equation}
The orthonormality of the modes are proven by calculating the inner products
\[
\left(  v_{\omega},v_{\omega^{\prime}}\right)  =i\int_{-\infty}^{\infty
}v_{\omega}(x)^{\ast}\partial_{x}v_{\omega^{\prime}}(x)dx.
\]
Through a change of coordinates $x^{\prime}=f(x)$, it is verified that
\[
\left(  v_{\omega},v_{\omega^{\prime}}\right)  =\left(  u_{\omega}%
,u_{\omega^{\prime}}\right)  =\delta\left(  \omega-\omega^{\prime}\right)  ,
\]
where
\begin{equation}
u_{\omega}(x)=\frac{1}{\sqrt{4\pi\omega}}e^{-i\omega x}. \label{09}%
\end{equation}
The completeness of $\left\{  v_{\omega}(x)\right\}  $ is trivial because
$v_{\omega}(x)$ has a one-to-one correspondence with $u_{\omega}(x)$ due to
the monotonicity of $f(x)$ and the fact that $\left\{  u_{\omega}(x)\right\}
$ is complete. Then, $\hat{\varphi}_{+}(x^{+})$ can be expanded in terms of
this mode function as
\begin{equation}
\hat{\varphi}_{+}(x^{+})=\int_{0}^{\infty}\left(  \hat{f}_{\omega}v_{\omega
}(x^{+})+\hat{f}_{\omega}^{\dag}v_{\omega}^{\ast}(x^{+})\right)  d\omega,
\label{07}%
\end{equation}
where $\hat{f}_{\omega}^{\dag}$ and $\hat{f}_{\omega}$ are creation and
annihilation operators, respectively, satisfying $\left[  \hat{f}_{\omega
},\hat{f}_{\omega^{\prime}}^{\dag}\right]  =\delta\left(  \omega
-\omega^{\prime}\right)  $ and $\left[  \hat{f}_{\omega},\hat{f}%
_{\omega^{\prime}}\right]  =0$ \cite{BD}. Let us introduce a quantum state
$|f\rangle$ such that%
\begin{equation}
\hat{f}_{\omega}|f\rangle=0 \label{06}%
\end{equation}
for all $\omega$. Because $v_{\omega}(x)$ is not a superposition of
$u_{\omega}(x)$ with positive frequency, the state $|f\rangle$ is a squeezed
vacuum state of the form
\[
|f\rangle\propto\exp\left(  \int_{0}^{\infty}d\omega\int_{0}^{\infty}%
d\omega^{\prime}\gamma_{\omega\omega^{\prime}}\hat{a}_{\omega}^{\dag}\hat
{a}_{\omega^{\prime}}^{\dag}\right)  |0\rangle,
\]
where $\hat{a}_{\omega}^{\dag}$ is a creation operator corresponding to
$u_{\omega}(x)$. For an arbitrary $|f\rangle$, the two-point correlation
function $\langle f|\hat{\Pi}_{+}(x)\hat{\Pi}_{+}(x^{\prime})|f\rangle$ is
evaluated as
\[
\langle f|\hat{\Pi}_{+}(x)\hat{\Pi}_{+}(x^{\prime})|f\rangle=-\frac
{\partial_{x}f(x)\partial_{x^{\prime}}f(x^{\prime})}{4\pi\left(
f(x)-f(x^{\prime})-i\epsilon\right)  ^{2}}.
\]
Thus, for $x_{A}\leq x_{2A}$ and $x_{A}^{\prime}\leq x_{2A}$,
\[
\langle f|\hat{\Pi}_{+}(x_{A})\hat{\Pi}_{+}(x_{A}^{\prime})|f\rangle
=\langle0|\hat{\Pi}_{+}(x_{A})\hat{\Pi}_{+}(x_{A}^{\prime})|0\rangle
\]
holds. Because $|f\rangle$ is a Gaussian state completely specified by
$\langle f|\hat{\Pi}_{+}(x)\hat{\Pi}_{+}(x^{\prime})|f\rangle$, the above
relation implies that the quantum fluctuation in $(-\infty,x_{2A}]$ is the
same as the zero-point fluctuation of $|0\rangle\langle0|$. This indicates
that $(-\infty,x_{2A}]$ is a local-vacuum-state region with zero energy. The
term \textquotedblleft zero energy" is used here to indicate not only that the
average value of the energy density in this region vanishes but also that all
correlations between the energy density operators are the same as those in the
vacuum state. Similarly, for $x_{B}\geq x_{1B}+T$ and $x_{B}^{\prime}\geq
x_{1B}+T$,
\[
\langle f|\hat{\Pi}_{+}(x_{B})\hat{\Pi}_{+}(x_{B}^{\prime})|f\rangle
=\langle0|\hat{\Pi}_{+}(x_{B})\hat{\Pi}_{+}(x_{B}^{\prime})|0\rangle
\]
holds, implying that $[x_{1B}+T,\infty)$ is also a local-vacuum-state region
with zero energy. After time $T$ has elapsed, the local-vacuum-state region
moves to $[x_{1B},\infty)$ because of the left-moving evolution.

Consider the case of a large $L$. At time $t=0$, Alice (who stays in the
zero-energy region of $(-\infty,x_{2A}]$) applies the measurements in
Eqs.~(\ref{11}) and (\ref{12}) to $\hat{\varphi}~$ in the state $|f\rangle$.
The measurement result $\mu$ is sent to Bob (who stays in the zero-energy
region of $[x_{1B},\infty)$) at $t=T$. During communication time $T$, the
information of $\mu$ jumps across the long-distance region $(x_{2A},x_{1B})$
with a positive finite energy $E_{C}$, evaluated as
\begin{equation}
E_{C}=\frac{1}{48\pi}\int_{x_{2A}}^{x_{1B}+T}\left(  \partial_{x}\ln\left(
\partial_{x}f(x)\right)  \right)  ^{2}dx. \label{08}%
\end{equation}
The excitation energy $E_{C}$ is so large that it can afford to maintain the
negative energy $-E_{B}$ generated by Bob because $E_{C}$ is placed near to
$-E_{B}$. Hence, the positive energy $E_{A}$ injected by Alice can be
separated far from $-E_{B}$, and a long-distance QET becomes possible. Since
$f(x)$~is a nonsingular $C^{1}$ function, $E_{C}$ is finite unless it exactly
attains $\partial_{x}f(x)=0$ with \thinspace$l=2d$. At $t=T$, Bob is able to
extract the teleported energy from $\hat{\varphi}$ by performing the same
operation as in Eq.~(\ref{13}) on the local zero-point fluctuation. Note that
\[
f(x_{B})-f(x_{A})=x_{B}-x_{A}-l.
\]
This means the effective distance for the correlation between the two points
in the state $|f\rangle$ is much less than the physical distance. By simply
replacing Eq.~(\ref{14}) with
\begin{equation}
\langle f|\hat{\Pi}_{+}(x_{B})\hat{\Pi}_{+}(x_{A})|f\rangle=-\frac{1}%
{4\pi\left(  x_{B}-x_{A}-l-i\epsilon\right)  ^{2}}, \label{9}%
\end{equation}
the amount of teleported energy $E_{Bf}$ can be evaluated as%

\begin{equation}
E_{Bf}=\frac{\left(  \int_{-\infty}^{\infty}\int_{-\infty}^{\infty}\frac
{g_{B}(x_{B})g_{A}(x_{A})}{\left(  x_{B}-x_{A}+T-l\right)  ^{3}}dx_{B}%
dx_{A}\right)  ^{2}}{\pi^{2}\int_{-\infty}^{\infty}\left(  \partial_{x}%
g_{B}(x)\right)  ^{2}dx\exp\left(  \frac{1}{\pi}\int_{0}^{\infty}\left\vert
\widetilde{g}_{A}(\omega)\right\vert ^{2}\omega d\omega\right)  }. \label{16}%
\end{equation}
The difference between Eq.~(\ref{3}) and Eq.~(\ref{16}) is just the appearance
of $l$ in the correlation function between the two separate regions of the
numerator integral. It should be noted that $l$ can, in principle, take a
large value satisfying $l\leq L+T$. By taking an $L$-dependent squeezed state
$|f\rangle$ such that
\begin{equation}
l\sim L+T, \label{17}%
\end{equation}
the long-distance damping of $E_{Bf}$ behaves not as $O(L^{-6})$ but as
$O(L^{0})$ as the distance $L$ increases. Therefore, this squeezed-state QET
indeed overcomes the distance bound in Eq.~(\ref{1}). Note that $E_{Bf}$ for
an $L$-independent $|f\rangle$ with a fixed $l$ again exhibits the original
damping behavior of $O(L^{-6})$ when $L\gg l$, as it should. The energy
distribution of the final state of the protocol is depicted in Fig.~1.

\begin{figure}[ptb]
\includegraphics[clip]{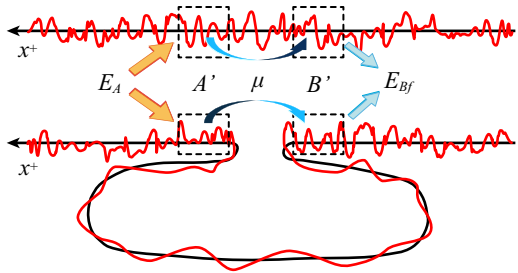}\caption{(Color online) Schematic showing
abrupt expansion of the space where quantum fluctuation of the field becomes
severely stretched. The upper figure depicts quantum fluctuation in the vacuum
state before the expansion, while the lower figure depicts the stretched
quantum fluctuation, which can be described by the modified mode function
$v_{\omega}(x^{+})$ after the expansion.}%
\end{figure}

The essence of QET without the limit of distance is as follows. The distance
dependence of $E_{B}$ for vacuum-state QET comes from $C_{\mu AB}$ as defined
in Eq.~(\ref{20}). This correlation is generated by the vacuum-state
entanglement between local zero-point fluctuations in the regions of Alice and
Bob. If we supply two local quantum fluctuations of $\hat{\varphi}$ that are
far away from each other and with the same entanglement and correlations as
those of the local zero-point fluctuations of two close regions, the QET
remains effective, independent of the distance, because $C_{\mu AB}$ is the
same. However, to sustain the negative-energy excitation generated by Bob,
additional positive energy of the field must be placed near the
negative-energy excitation. In this new protocol, the negative energy is
sustained by the positive energy of the squeezed region. As depicted in
Fig.~2, such a long distance correlation is indeed realized by an abrupt
expansion of the space over which the quantum field $\hat{\varphi}$ exists,
analogous to cosmological inflation in general relativity. The upper diagram
in Fig.~2 depicts the vacuum fluctuation of $\hat{\varphi}$ in the
state-preparation region, which is equipped on the right-hand side of $B$,
before moving left to the original QET experimental region of $A$ and $B$.
Assume that, if we perform the vacuum QET from $A^{\prime}$ to $B^{\prime}$,
the distance between $A^{\prime}$ and $B^{\prime}$ in Fig.~2 is so small that
the amount of teleported energy attains a large value of $E_{Bf}$. The lower
diagram in Fig.~2 shows the sudden expansion of the small subspace between
$A^{\prime}$ and $B^{\prime}$, which generates local excitation of
$\hat{\varphi}$ due to severe stretching of the field modes. Note that the
stretched mode function can be described by $v_{\omega}(x^{+})$ in a similar
way to expanding Universe models including cosmological inflation
\cite{KT,BD}. The expansion is specified by the metric
\begin{equation}
ds^{2}=g_{\mu\nu}dx^{\mu}dx^{\nu}=dt^{2}-\left(  a(t,x)dx\right)  ^{2},
\label{110}%
\end{equation}
the scale factor of which obeys $a(t,x)=1$ outside the stretching region. Let
us assume that the expansion is very rapid so that we can regard it
instantaneous, and the expansion happens at time $t=0$. The initial condition
of the scale factor is $a(t=-0,x)=1$, and the equation of motion of
$\hat{\varphi}$ in the expansion is given by
\begin{equation}
\partial_{\mu}\left(  \sqrt{-g}g^{\mu\nu}\partial_{\nu}\hat{\varphi}\right)
=0. \label{ceq}%
\end{equation}
When we consider the plane-wave mode function $u_{\omega}(x)$ in
Eq.~(\ref{09}) just before the rapid expansion, the form of the mode function
remains unchanged under the instantaneous expansion of space at the coordinate
$(t,x)$. However, from Eq.~(\ref{110}), the correct physical distance
coordinate $X$ should be obtained by
\[
X=X(x)=\int_{0}^{x}a(+0,x^{\prime})dx^{\prime}.
\]
Thus, the mode function after the expansion is computed in terms of the
physical coordinate $(t,X)$ as
\[
v_{\omega}(X)=\frac{1}{\sqrt{4\pi\omega}}e^{-i\omega x(X)},
\]
where $x(X)$ is the inverse function of $X(x)$. This indeed reproduces the
squeezed mode function in Eq.~(\ref{05}) by defining $f(X)=x(X)$. Thus, $f(x)$
in Eq.~(\ref{05}) is computed from the relation
\[
f^{-1}(x)=\int_{0}^{x}a(+0,x^{\prime})dx^{\prime},
\]
where $f^{-1}$ is the inverse function of $f$. Hence, the quantum state after
the expansion is equivalent to the squeezed vacuum state given by
Eq.~(\ref{06}). Since the field is still in a local vacuum state outside the
expanded region, the correlation between $A^{\prime}$ and $B^{\prime}$ remains
unchanged. Hence, if QET from $A^{\prime}$ to $B^{\prime}$ is executed, we
would be able to teleport the same amount of energy $E_{Bf}$, although the
distance would become very large. After the sudden expansion, the long-range
correlated quantum fluctuation moves left into the experimental region
including $A$ and $B$ and can be used for the original long-distance QET from
$A$ to $B$ with teleported energy $E_{Bf}$.

\begin{figure}[ptb]
\includegraphics[clip]{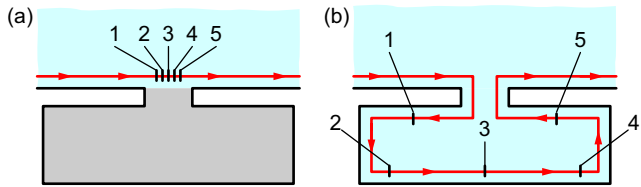}\caption{(Color online) Schematics of quantum
Hall edge current (red lines) (a) before and (b) after the expansion.}%
\end{figure}

Such spatial expansion, for example, may be performed in quantum Hall edge
current systems \cite{yih}. Recall that the current [red line and arrows in
Fig. 3(a)] is well described by a chiral (unidirectional) massless field in
one dimension \cite{yoshioka}. A local extrusion of bulk electrons toward the
outside [Fig. 3(b)] can be experimentally obtained by dynamically controlling
the electron density in the depleted region [the gray region in Fig. 3]. Such
operations are commonly used in field-effect transistors at frequencies up to
the subterahertz regime. Since marked points 1--5 on the current trajectory in
Fig. 3(a) are separated as in Fig. 3(b) after the extrusion, the transition
describes an expansion of space in which the massless field exists. The field
satisfies Eq.~(\ref{ceq}) in terms of a metric $g_{\mu\nu}$ induced by the
extrusion, creating a continuously parameterized multi-mode squeezed state.
Generation of such a squeezed state is well known in research of quantum
fields in curved spacetime, such as inflationary universes \cite{KT}, and
provides an interesting squeezing method even in condensed matter physics.
Finally, it should be noted that high-precision squeezed-state generation is
not required for long-distance QET. However, it is imperative that a very long
detour path with length $l$ for the edge current be inserted between two
very-close local vacuum regions just as $A^{\prime}$ and $B^{\prime}$ in Fig.
2. This achieves the same amount of $E_{B}$ in Eq.~(\ref{16}). This spatial
expansion method is one strategy for realizing long-distance correlation, thus
facilitating experimental verification of QET and potentially contributing to
quantum device applications.


\section{\bigskip Summary}

In this paper, we have pointed out that vacuum-state QET suffers from the
distance bound in Eq.~(\ref{1}). The bound is a severe obstacle to the
implementation of long-distance QET in nanophysics. To overcome the bound on
the distance $L$ between the sender and receiver of QET, we proposed a new QET
protocol that adopts a squeezed vacuum state defined by Eq.~(\ref{06}), which
corresponds to the mode function in Eq.~(\ref{05}) with a $C^{1}$ function
$f(x)$ defined by Eqs.~(\ref{01})--(\ref{04}). The measurement of Alice and
the local operation of Bob are the same as those of the vacuum-state QET and
are given by Eqs.~(\ref{11}), (\ref{12}), and (\ref{13}). By taking an
$L$-dependent squeezed state $|f\rangle$ such that the parameter $l$ of $f(x)$
satisfies Eq.~(\ref{17}), the long-distance damping of $E_{Bf}$ behaves as
$O(L^{0})$. Therefore, QET without the limit of distance can be attained.
Long-distance QET may be experimentally verified by adopting a spatial
expansion method in quantum Hall edge currents, which is a new scheme to
create a continuously parameterized multi-mode squeezed state in condensed
matter physics.

~

\textbf{Acknowledgments}\newline

\begin{acknowledgments}
G. Y. is supported by a Grant-in-Aid for Scientific Research (No. 24241039)
from the Ministry of Education, Culture, Sports, Science and Technology
(MEXT), Japan.
\end{acknowledgments}


\begin{thebibliography}{99}
\bibitem {m}For example, a general measurement using a nanoscale RC circuit
has been proposed as a realistic QET experiment in quantum Hall edge currents
\cite{yih}.
\end{thebibliography}
\end{document}